# A high-performance all-silicon photodetector enabling telecom-wavelength detection at room temperature

Mohd Saif Shaikh,[1,2*] Mircea-Traian Cătuneanu,[2] Ahmad Echresh,[1] Rang Li,[1,2] Shuyu Wen,[1] Guillermo Godoy-Pérez,[1,2] Slawomir Prucnal,[1] Manfred Helm,[1,2] Yordan M. Georgiev,[1,3] Kambiz Jamshidi,[2] Shengqiang Zhou,[1**] & Yonder Berencén,[1***]

[1]*Helmholtz-Zentrum Dresden-Rossendorf, Institute of Ion Beam Physics and Materials Research, Bautzner Landstrasse 400, 01328 Dresden, Germany*

[2]*Dresden University of Technology, 01062 Dresden, Germany*

[3]*Acad. E. Djakov Institute of Electronics, Bulgarian Academy of Sciences, 72 Tzarigradsko Chaussee blvd., 1784 Sofia, Bulgaria*

Corresponding authors: *m.shaikh@hzdr.de**; *s.zhou@hzdr.de***; *y.berencen@hzdr.de****

**Photonic integrated circuits (PICs) are crucial for advancing optical communications, promising substantial gains in data transmission speed, bandwidth, and energy efficiency compared to conventional electronics[1]. Telecom-wavelength photodetectors, operating near 1550 nm, are indispensable in PICs, where they enable the sensitive and low-noise conversion of optical signals to electrical signals for efficient data processing. While silicon is ideal for passive optical components, its limited absorption in the optical telecommunication range (1260-1625 nm) typically necessitates integrating an alternative material, such as germanium[2], for photodetection — a process that introduces significant fabrication challenges[3]. Here, we present a high-performance, all-silicon waveguide-coupled photodetector, which operates at**

**room temperature within the optical telecom C band. By introducing deep-level impurities into silicon at concentrations close to the solid-solubility limit, we enable efficient sub-bandgap absorption without compromising recombination carrier lifetimes and mobilities. This detector achieves a responsivity of 0.56 A/W, a quantum efficiency of 44.8%, a bandwidth of 2 GHz, and a noise-equivalent power of $4.2\times10^{-10}$ W/Hz$^{1/2}$ at 1550 nm, fulfilling requirements for telecom applications. Our approach provides a scalable and cost-effective solution for the monolithic integration of telecom-wavelength photodetectors into silicon-based PICs, advancing the development of compact photonic systems for modern communication infrastructures.**

The escalating demands for bandwidth, speed, and energy efficiency in modern computing systems drive the adoption of optical interconnects, which offer several advantages over conventional electronic links, including lower latency, higher data transmission rates, and reduced power consumption[1]. These benefits have made photonic integrated circuits (PICs) a promising foundation for next-generation communications and data processing, where dense, efficient, and scalable optical components are essential to meet the demands of high-performance computing and AI-driven applications[4,5]. Among these components, telecom-wavelength photodetectors (PDs) are indispensable, as they convert optical signals into electrical signals for communication and data processing[6].

Silicon photonics, leveraging CMOS-compatible processes, enables the development of cost-efficient, high-density PICs that seamlessly integrate optical I/O modules with conventional electronic processors[1,7]. While silicon's transparency in most of the optical telecommunication bands makes it well-suited for passive optical components like waveguides, couplers, and splitters, its bandgap (~1100 nm cutoff) precludes efficient detection at telecom wavelengths around 1550 nm. To address this demand, photodetection has traditionally relied on heteroepitaxial integration of germanium (Ge) with silicon[2], enabling responsivity at telecom wavelengths. However, Ge-based photodetectors present inherent challenges, including high dark currents, thermal sensitivity, and costly additional growth processes that complicate scaling and limit compatibility with CMOS foundries.

Extensive research has sought to overcome these limitations through alternative silicon-based sub-bandgap photon detection approaches. Hyperdoping of silicon with deep-level impurities at concentrations well above their equilibrium solubility limit provides a pathway to achieve room-temperature sub-bandgap absorption for telecom wavelengths while maintaining CMOS compatibility[8,9]. Alternatively, techniques such as two-photon absorption (TPA), photon-assisted tunneling (PAT), the creation of dislocation loops, and optical resonant enhancement in microring resonators (MRRs) have also shown promise in enhancing responsivity and absorption efficiency at telecommunication wavelengths[10–12]. However, each approach faces trade-offs; for instance, TPA and PAT suffer from inherently low responsivity, dislocation loops enable only weak infrared absorption, and resonant cavities like MRRs, while enhancing absorption of TPA and PAT through resonant light trapping and field amplification, introduce crosstalk in densely packed wavelength-division multiplexing systems and limit bandwidth due to long photon lifetimes. Additionally, hyperdoping strategies that push impurity concentrations above the solid-solubility limit result in extremely short carrier recombination lifetimes (1-200 picoseconds)[13,14] and significantly reduced carrier mobilities (10-40 $cm^2/V.s$)[13,15]. Although this approach enhances room-temperature photoresponse at telecommunication wavelengths[16,17], it compromises key performance metrics of the photodetector[13].

In this work, we demonstrate a scalable, high-performance solution: an all-silicon waveguide-coupled p-i-n photodetector operating at telecom wavelengths, enabled by introducing Te impurities at concentrations close to the equilibrium solid-solubility limit ($3.5\times10^{16}$ $cm^{-3}$)[18] to create deep-level intragap states. Tellurium doping can be made CMOS-compatible through localized implantation and back-end-of-line integration, using controlled thermal budgets and dedicated processing tools to prevent contamination and evaporation. This method allows room-temperature sub-bandgap photoresponse across all the optical telecommunication bands while avoiding the issues of high carrier recombination and low carrier mobility associated with hyperdoped silicon approaches. Operating within the optical C band, our detector achieves a responsivity of 0.56 A/W, a quantum efficiency of 44.8%, a noise-equivalent power (NEP) of $4.2\times10^{-10}$ $W/Hz^{1/2}$, a linear dynamic range of 33.7 dB, and a bandwidth of 2 GHz. This CMOS-compatible approach significantly simplifies fabrication, reduces costs, and supports monolithic integration within silicon photonic systems. Our results highlight the potential of Te-implanted Si photodetectors as a viable alternative to hybrid Ge-based solutions, advancing the integration of efficient, compact, and scalable photonic components.

## Results

### Device structure and fabrication

In a photonic-grade silicon-on-insulator (SOI) chip with a 220-nm-thick Si device layer over a 2000-nm buried oxide (BOX) layer, we designed and fabricated monolithic waveguide-coupled Si p-i-n photodetectors (WG-PDs) with varying detector absorption lengths (L) ranging from 300 μm to 1000 μm (Fig. 1a). Light is coupled into the chip via focusing grating couplers, each 25 μm wide and 20 μm long, which are connected to a 200-μm-long adiabatic taper that transitions into a single-mode rib waveguide. The waveguide segment between the taper and the detector is 30 μm long and is identical across all devices. Grating couplers and rib waveguides were designed and optimized using finite-difference time-domain (FDTD) simulations[19] to couple off-chip light from a 1550 nm single-mode fiber laser into the photodetector, which then converts the incident light into a measurable photocurrent (Fig. 1b).

Grating couplers, along with tapers, are used to couple and confine the 1550 nm light from the fiber laser into a single-mode rib waveguide, with the light incident at a 10-degree angle relative to the surface normal of the device layer. The p-i-n photodetector is fabricated at the end of the single-mode rib waveguide (Fig. 1c).

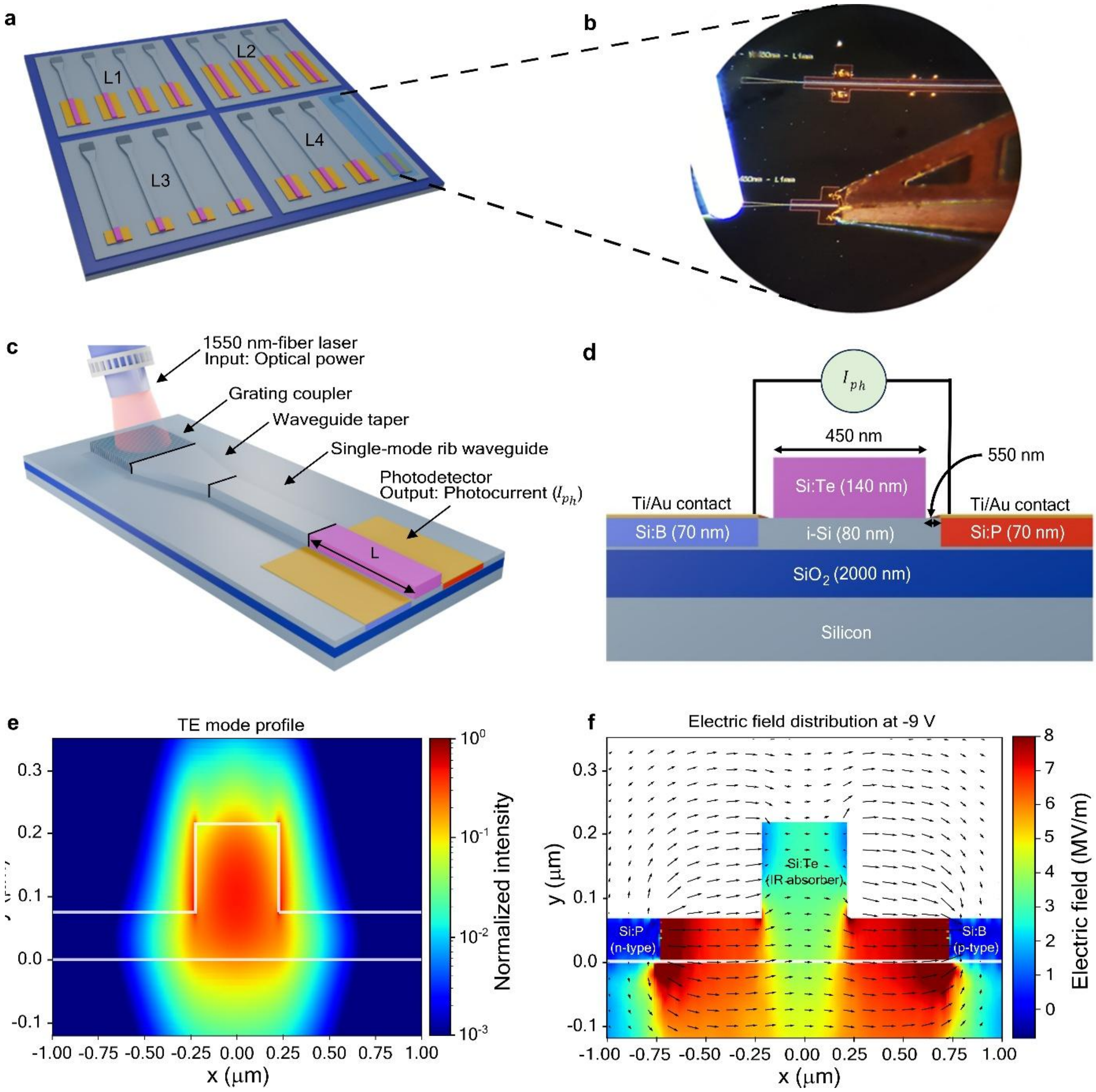

**Figure 1 | Design of the waveguide-coupled Si:Te p-i-n photodetector. a,** Schematic of an SOI chip with photodetectors, each featuring different absorption region lengths (L1, L2, L3, and L4) coupled to waveguides and grating couplers. **b,** Optical microscope image of a WG-PD, showing the 1550 nm-fiber laser (left) used to couple light through the grating coupler and detect a measurable photocurrent in the detector with electrical probes (right). **c,** Schematic of the full WG-PD. Incident light from the 1550 nm-fiber laser is coupled into the SOI chip through a grating coupler and then confined into a single-mode rib waveguide via a waveguide taper. The waveguide is laterally coupled to the detector, where the light is detected by the p-i-n junction, producing a measurable photocurrent. **d,** Cross-sectional sketch of the p-i-n WG-PD, illustrating the Te-implanted Si optical absorption region optimized for telecom wavelengths, integrated along the single-mode rib waveguide. Below this, the intrinsic region is sandwiched between B-doped (p-type) and P-doped (n-type) Si regions, with Ti/Au metal contacts simultaneously formed on the p- and n-doped regions. All schematics are not drawn to scale. **e,** Simulated transverse electric (TE) mode profile in a cross-sectional view of the waveguide at the operating wavelength of 1550 nm, showing strong field confinement in the Si:Te absorption region. The color scale represents the normalized optical intensity on a logarithmic scale. **f,** Cross-sectional simulation of the magnitude of the total electric field used to illustrate the spatial distribution and confinement of the electric field within the device at an applied reverse bias of -9 V. The electric field in the Si:Te region facilitates in the separation of photogenerated carriers.

The cross-section of the detector's active area (140 nm etch depth and 450 nm width) is created by implanting deep-level Te impurities at two different energies through a window, forming a 140 nm-deep region with uniform Te doping. Next, p- and n-type regions are formed by implanting B and P, respectively, through two additional windows on the chip floor, each positioned 550 nm away from the edges of the Te-implanted region (Fig. 1d). The P and B dopants were first activated using a rapid thermal annealing (RTA) process, followed by a separate RTA step with distinct parameters to activate the Te atoms. Subsequently, Ti/Au metal contacts were formed on the p- and n-doped regions. Details of the fabrication process are provided in Methods and the Supplementary Information, figures S1 - S6.

Figure 1e and f illustrate the simulated transverse electric (TE) mode profile and electric field distribution of the Si:Te photodetector in cross-section. The TE mode profile in Figure 1e shows strong optical field confinement within the Si:Te absorption region, ensuring efficient coupling of the guided mode into the detector. The intensity distribution reveals that most optical power remains concentrated in the upper portion of the rib waveguide, where deep-level impurities in Si:Te enable sub-bandgap absorption. The intrinsic Si layer underneath the Si:Te region remains un-implanted to serve as a high-mobility drift region, free of implantation-induced traps. This design minimizes carrier recombination and trap-assisted losses, facilitating efficient photocarrier transport and collection, thus enhancing responsivity and reducing 1/f noise.

Figure 1f shows the magnitude of the total electric field in the cross-section, revealing its spatial distribution within the Si:Te region under reverse bias of -9 V. The primary transport occurs along the horizontal direction (x-axis). The high electric field (~3-4 MV/m) in the Si:Te region drives efficient carrier separation, minimizing recombination losses. The photogenerated carriers are then swept through the high-field intrinsic region to the p- and n-doped regions and collected at the contacts.

**Device characterization**

Figure 2a shows a dark current-voltage (I-V) curve for the WG-PD at room temperature under both forward and reverse bias conditions. In reverse bias, the WG-PD exhibits rectifying behavior, with a dark current of about $1.3\times10^{-8}$ A at a reverse bias of -9 V.

Before conducting the optoelectronic characterization of the WG-PD, we deliberately measured over 20 control passive devices that included an input grating coupler, a single-mode rib waveguide, and an output coupler. This was done to estimate the total optical loss from these structures, which must be considered when calculating responsivity. Our analysis revealed that the average optical loss is approximately 10 dB, with the majority of this loss attributed to the taper and grating couplers (details in the Supplementary Information, figures S2, S3, S7 and S8). The waveguide and absorbing region form a continuous Si rib waveguide, with the Te-implanted region confined to the center of the rib. This avoids abrupt refractive index contrast, minimizing mode mismatch and reflection. The effective mode index (~2.3 at 1550 nm) remains nearly uniform along the propagation direction, ensuring low back-reflection.

The responsivity, defined as $\mathcal{R} = I_{ph}/P_{in}$, where $I_{ph}$ is the photocurrent in amperes and $P_{in}$ is the waveguide-coupled optical power in watts incident on the WG-PD, is measured as a function of input optical power with an excitation wavelength of 1550 nm and applying a reverse bias of -9 V across electrodes (Fig. 2b). Notably, the reported responsivity is calculated with respect to the on-chip optical power—that is, after accounting for the measured grating coupler loss—thereby reflecting the relevant figure of merit for integrated photonic applications. We observed a decrease in on-chip responsivity from 0.56 A/W to 0.13 A/W at higher input optical powers. This reduction may be attributed to several interconnected factors: i) elevated carrier densities can enhance recombination processes, particularly via non-radiative pathways, thereby reducing the net photocurrent; and ii) Auger recombination, which becomes increasingly significant at high carrier concentrations, may further contribute to the loss in responsivity. However, the precise origin of the saturation effect remains to be elucidated and is beyond the scope of this study.

We found that the responsivity at 1550 nm under an input optical power of 125 nW incident on the WG-PDs increases from 0.12 A/W to 0.56 A/W as the reverse bias increases (Fig. 2c). This increase is primarily due to the higher electric field in the Te-implanted Si region (see Fig. 1f), which accelerates the separation and collection of photogenerated carriers. The stronger field enhances the drift velocity of electrons and holes, reducing recombination losses and thereby boosting both photocurrent and responsivity. Additionally, a higher reverse bias expands the depletion region, allowing more incident light to be absorbed where carriers are swiftly collected, further enhancing efficiency. In contrast, at zero bias, responsivity is three orders of magnitude lower because carrier separation depends solely on the slower diffusion process without an assisting electric field, leading to higher carrier recombination rates and thus lower responsivity.

We calculate the quantum efficiency ($\eta$) of the photodetectors at 1550 nm under a reverse bias of -9 V as $\eta(\lambda) = \mathcal{R}_\lambda hc/\lambda e$, where $\lambda$ is the wavelength (nm), $\mathcal{R}_\lambda$ is photodetector responsivity (A/W) at a particular wavelength ($\lambda$), $h$ is the Planck constant, $c$ is the light speed in vacuum, and $e$ is the elementary charge. The quantum efficiency was found to be $\eta$=44.8%.

Next, we characterized the WG-PDs for NEP, which is the minimum incident optical power that a photodetector can distinguish from noise, indicating its sensitivity. Further details can be found in the Methods and in Section 3a of the Supplementary Information.

In Fig. 2d, we present the NEP at 1550 nm measured under a reverse bias of -4 V, which was found to be $4.2\times10^{-10}$ W/(Hz)$^{1/2}$. This value is two orders of magnitude higher than commercial Ge photodetectors[20]. Despite the higher NEP, Te-implanted Si detectors offer key advantages for silicon photonic circuits, particularly in telecom-band applications, where reduced fabrication complexity and cost, integration, and scalability are prioritized over ultra-low NEP[21]. A potential strategy to lower the NEP in the Te-implanted WG-PDs involves optimizing the doping concentration and thickness of the Te-implanted layer to enhance sub-bandgap absorption efficiency while minimizing dark current.

The linear dynamic range ($\mathrm{LDR}$), defined as $\mathrm{LDR\,(dB)} = 10 \times \log_{10}(\mathrm{P_{max}}/\mathrm{P_{min}})$, represents the range of input optical power levels in watts, from the minimum ($\mathrm{P_{min}}$) to the maximum ($\mathrm{P_{max}}$), over which the detector exhibits a linear response to incident light. In this work, the $\mathrm{LDR}$ was found to be 33.7 dB for optical power levels ranging from $4.2\times10^{-10}$ to $1\times10^{-6}$ W, demonstrating the detector's capability to detect a wide range of optical power levels while maintaining linearity in its response.

The WG-PDs with longer optical absorption lengths maintain similar responsivity at 1550 nm, yet they exhibit an order of magnitude higher NEP and slightly increased capacitance (details in the Supplementary Information, Table S1). The consistent responsivity across varying device lengths ($L$) suggests that the 1550 nm laser light is fully absorbed within a 300 μm path, indicating effective optical absorption over this distance.

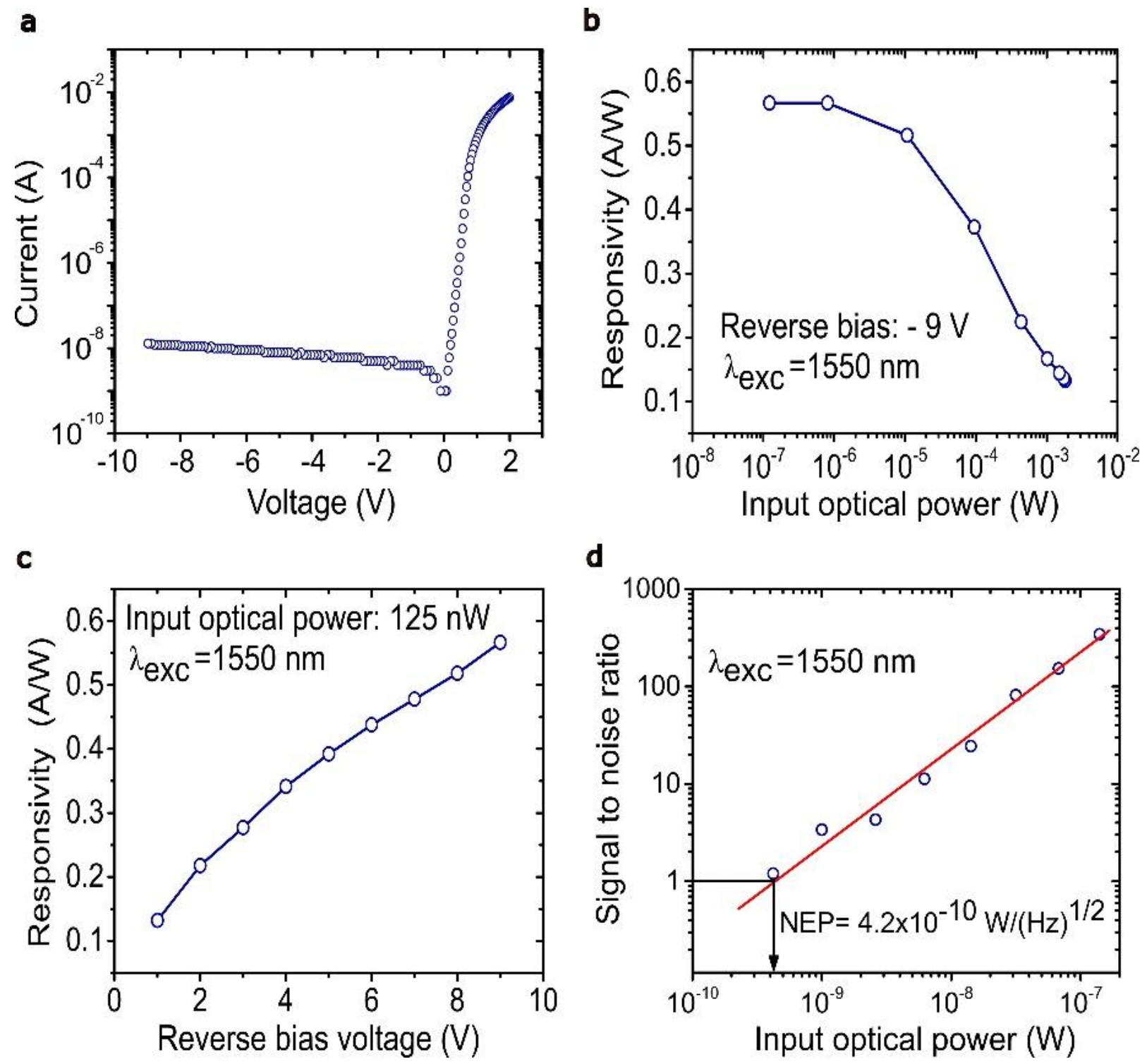

**Figure 2 | Room-temperature performance characteristics of an all-silicon WG-PD with a Si:Te absorption length of L=300 μm. a,** Dark current-voltage (I-V) characteristic under forward and reverse bias. **b,** Responsivity of the WG-PD biased at -9 V as a function of input optical power at a wavelength of 1550 nm. **c,** Responsivity at 1550 nm for an input optical power of 125 nW, plotted against the reverse bias. **d,** Signal-to-noise ratio (SNR) derived from frequency-dependent current measurements as a function of input optical power at 1550 nm for the WG-PD at reverse bias of 4 V, used to accurately determine the noise-equivalent power (NEP). The responsivity values in b and c include an overall optical loss (i.e. both coupling and propagation losses) of 10 dB, determined from over 20 grating-waveguide-grating control structures with identical dimensions to the WG-PDs.

Next, we determined the bandwidth ($f_{BW} = 1/2\pi R_{Total} C_j$) and the corresponding rise time response ($t_r = 0.35/f_{BW}$) of the WG-PDs using the measured junction capacitance ($C_j$= 5.4×10$^{-13}$ F) obtained from the C-V characteristic (Fig. 3a) and the total resistance $R_{Total} = R_L + R_s$= 179 Ω, where $R_L$=50 Ω is the load resistance and $R_s$= 129 Ω is the series resistance determined from the forward-bias I-V fitting to a single-diode model. The resulting theoretical bandwidth is approximately 1.6 GHz, corresponding to a rise time of ~219 ps. This theoretical value shows good agreement with the experimentally determined -3dB cutoff frequency of 2 GHz obtained from the frequency response measurement (Fig. 3b), validating the RC-limited model and highlighting the photodetector's potential for high-speed applications in integrated photonic circuits.

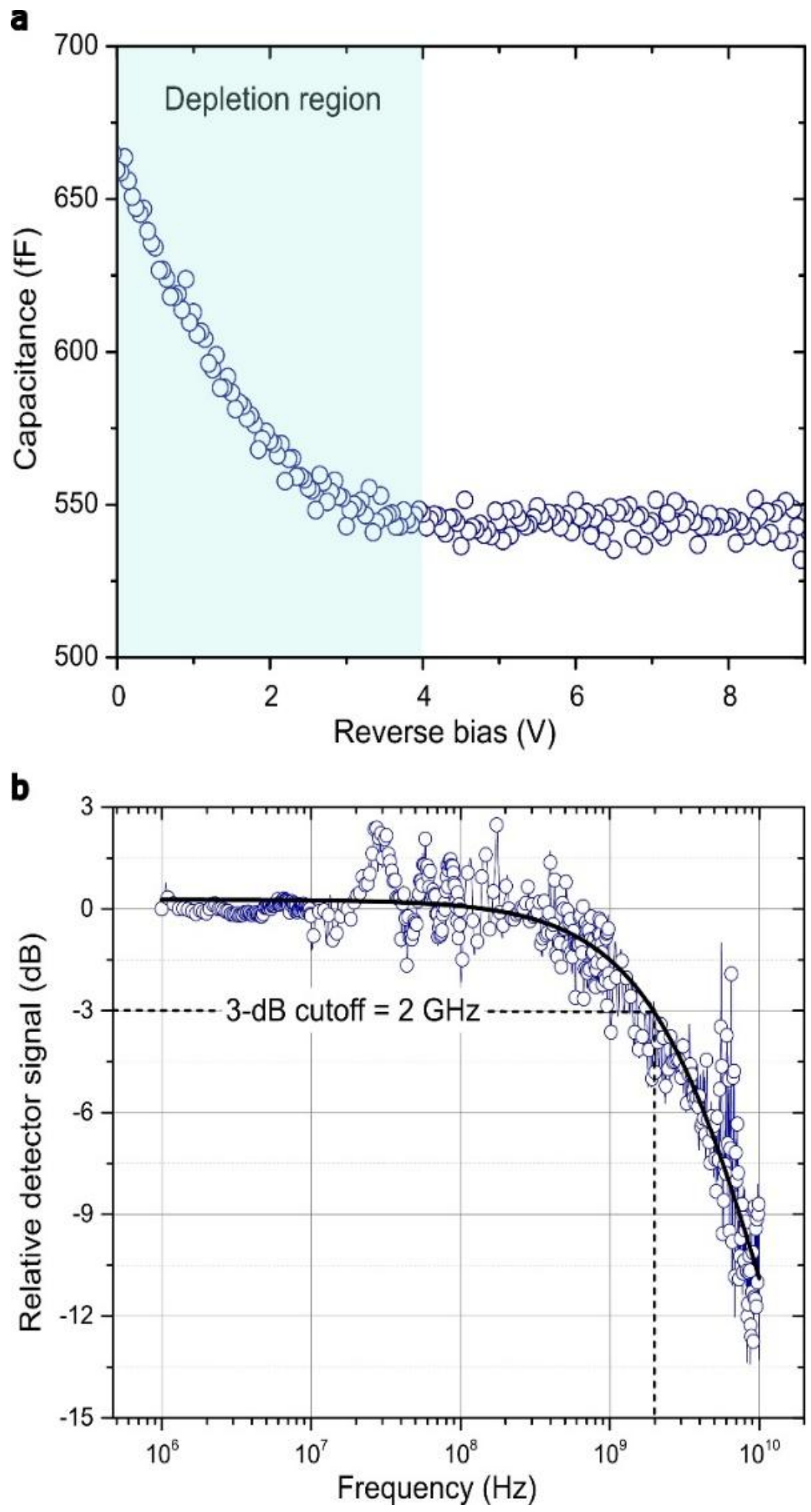

**Figure 3 | Frequency response characteristic of the all-silicon WG-PD at 1550 nm. a,** Capacitance–voltage (C-V) characteristics of the device measured at 1 MHz, showing a sharp decrease in capacitance with increasing reverse bias as the depletion region forms and expands. The shaded region indicates the onset of full depletion at a reverse bias of 4 V. **b,** Frequency response of the photodetector under reverse bias of

-9 V, showing a -3 dB cutoff frequency at 2 GHz. The black line is a fit to the experimental data (blue circles), indicating the high-speed operation of the device.

**Working principle of the telecom-wavelength p-i-n waveguide-coupled Si:Te photodetector**

The p-i-n waveguide-coupled Si:Te photodetector leverages sub-bandgap absorption enabled by deep-level Te impurities introduced into the silicon lattice. Acting as double donors, Te atoms create discrete energy levels approximately 0.2–0.4 eV below the conduction band edge[17]. These mid-gap states facilitate efficient absorption of telecom-wavelength photons (~1550 nm, 0.8 eV), where a photon can promote an electron either from the valence band to the Te impurity level (VB → $Te^{o}$), or directly from the Te levels into the conduction band ($Te^{o}$ → CB), depending on the photon energy. In the VB → $Te^{o}$ case, the photoexcited electron becomes temporarily trapped in the Te state, while a hole is generated in the valence band and swept toward the p-contact by the built-in electric field, contributing to the photocurrent.

Although the electron is initially localized in the deep Te level, it may subsequently be released into the conduction band via electric-field-assisted tunneling, consistent with the Poole–Frenkel effect. The internal electric field in the intrinsic region lowers the Coulomb barrier of the trap, enabling tunneling of the bound electron into the conduction band. Thus, this electron is rapidly swept toward the n-contact, further contributing to the photocurrent. This process, driven by sub-bandgap photon absorption and field-assisted tunneling, provides an efficient mechanism for photodetection at telecom wavelengths using a monolithic silicon platform. The energy band diagram illustrating the photoresponse mechanism is provided in Supplementary Fig. S12.

To investigate the nature of the photoexcitation mechanism, we measured the photocurrent as a function of incident optical power on a log-log scale. The data exhibit a sublinear dependence with a power-law exponent of approximately 0.82 (see Supplementary Fig. S13), consistent with a sequential two-photon absorption process involving the Te impurity level. This behavior supports the proposed mechanism in which the first photon excites an electron from the valence band to the Te level, followed by a second photon promoting it to the conduction band. The observed *two-step* absorption proceeds via a real Te impurity state, with sublinear scaling resulting from saturation of the finite intermediate-state population, unlike the quadratic dependence characteristic of simultaneous two-photon absorption.

**Discussion**

Next, we highlight the unique performance characteristics of Te-implanted Si p-i-n photodetectors operated at near-equilibrium Te concentrations, particularly in waveguide-coupled geometries. This configuration positions them as a promising candidate for very-large-scale integration (VLSI) into PICs via monolithic fabrication. Unlike hyperdoped Si, where impurity concentrations exceed the solid solubility limit, leading to heavy n-type doping, metallic-like behavior, and elevated dark currents and noise despite enhanced infrared absorption[16,22], our approach employs Te concentrations close to the solubility threshold. This strategy enables sub-bandgap photon absorption without introducing significant thermally generated free carriers, preserving low dark current operation. As a result, these devices exhibit a compelling trade-off between performance and process simplicity, achieving responsivities comparable to those of Ge photodetectors (see Table 1), while eliminating the need for non-conventional laser or flash lamp annealing. Crucially, the entire fabrication process remains fully compatible with standard CMOS technology.

Waveguide-coupled detectors exhibit substantial advantages over traditional vertical detector designs by enabling extended optical absorption lengths along the rib waveguide structure, reaching hundreds of microns, thus significantly improving responsivity. When implemented on SOI substrates, these devices achieve substantial noise reduction, as the photodetectors are isolated from the bulk Si, eliminating noise-inducing interactions and enhancing signal integrity. Additionally, waveguide-coupled Te-implanted Si detectors can be fabricated with compact footprints, lowering their capacitance and enhancing speed, which is a critical advantage for achieving high bandwidths and fast response times.

Recently, double microring resonators (d-MRRs) were demonstrated as an effective solution to address the issue of crosstalk in wavelength-division multiplexing systems[23] by significantly reducing spectral tails around resonances, a limitation observed in single-MRR designs. Their ability to suppress crosstalk while maintaining high channel capacity makes them an ideal choice for integration with Te-implanted Si photodetectors. The combination of d-MRRs and Te-implanted Si enables enhanced absorption and responsivity, leveraging the material's high absorption coefficient and the resonant enhancement of the MRRs.

**Table 1 | State-of-the-art performance of integrated photodetectors in silicon operating at 1550 nm.**

| Detector type | Material platform | Responsivity (A/W) | Dark current | QE (%) | NEP ($W/Hz^{1/2}$) | Bandwidth (GHz) | CMOS fab readiness |
|---|---|---|---|---|---|---|---|
| This work | All-silicon | 0.56 | 13 nA | 44.8 | $4.2\times10^{-10}$ | 2 | Production-ready[a] |
| Waveguide-coupled[2] | Si/Ge | 0.30 | 200 nA | - | - | 265 | Pilot/early production (some fabs)[b] |
| Double-microring resonator (MRR)[23] | All-silicon | 0.40 | 1 nA | - | - | 40 | Production-ready[a] |
| Waveguide-coupled[24] | Si/III-V | 0.20 | 60 pA | - | - | 50 | R&D to pilot line[c] |
| Waveguide-integrated[25] | Si/Graphene | 0.65 | - | 52 | $4.6\times10^{-11}$ | 65 | R&D only[d] |

[a]**Production-ready**: Full support in commercial CMOS lines. High-volume capable. [b]**Pilot/early production**: Available in select lines or under restricted conditions. Moderate yield. [c]**R&D to pilot line**: Active development in specialized labs or semi-commercial pilot lines. [d]**R&D only**: No mainstream foundry support. Custom or academic processing only.

In summary, we have demonstrated a high-performance, monolithically integrated, all-silicon photodetector that operates efficiently within the optical telecommunication C band at room temperature. By engineering deep-level impurity states at concentrations approaching the solid-solubility limit, we achieve strong sub-bandgap absorption in silicon while preserving favorable electronic transport properties. The resulting device exhibits a responsivity of 0.56 A/W, a quantum efficiency of 44.8%, a 3 dB bandwidth of 2 GHz, and a noise-equivalent power of $4.2\times10^{-10}$ W/Hz$^{1/2}$ at 1550 nm, addressing key performance metrics required for optical communication systems. While the 2 GHz bandwidth of our Si:Te waveguide-coupled photodetector is modest compared to that of ultra-fast devices, its high on-chip responsivity, high quantum efficiency, low dark current, and monolithic compatibility with commercial CMOS processes make it a compelling platform for scalable photonic integration at telecom wavelengths.

As summarized in Table 1, devices relying on III-V materials or graphene achieve high bandwidths, but typically require non-standard processes or materials not supported in production CMOS lines, limiting their scalability and integration. By contrast, our all-silicon device operates in the telecom range while remaining entirely within CMOS-compatible process flows, marking a critical step toward large-scale integration.

These results establish a new paradigm for photodetection in silicon photonic integrated circuits, eliminating the need for heterogeneous integration of alternative materials such as germanium. The compatibility of this approach with standard silicon processing techniques ensures scalability, cost-effectiveness, and CMOS co-integration. Beyond telecommunications, our platform may also enable advances in emerging fields such as mid-infrared sensing, and on-chip optical interconnects, where CMOS-compatible detection is critical.

## Methods

### *Fabrication*

SOI substrates with a 220 nm intrinsic silicon layer (resistivity: 14-18.9 Ω.cm) and a 2000 nm buried $SiO_2$ layer were utilized to fabricate on-chip waveguide-coupled silicon p-i-n photodetectors using electron beam lithography (EBL) and inductively coupled plasma reactive ion etching (ICP-RIE). The fabrication process included cleaning the substrates, applying a negative tone resist (hydrogen silsesquioxane), and exposing the patterns via EBL. After development, the patterns were transferred into the silicon layer through ICP-RIE. To form a p-i-n junction for efficient extraction of photogenerated electron-hole pairs, phosphorus and boron dopants were implanted to create n- and p-type regions, respectively. These regions were then subjected to annealing to activate the dopants and repair the implantation-induced defects. Additionally, tellurium ions were locally implanted to enable sub-bandgap absorption in silicon, with a rapid thermal annealing process conducted to repair implantation-induced defects and activate the Te dopants (see Supplementary Information for further details, Fig. S5 and S6).

### *Optical mode simulation*

We simulated the fundamental transverse electric (TE) optical mode of the waveguide using the MODE solver in Ansys Lumerical at a wavelength of 1550 nm. We chose the TE mode due to the high TE selectivity of the input grating coupler. To evaluate the optical absorption profile, we overlapped the normalized optical mode intensity distribution with the spatial extent of the tellurium-implanted region. We then computed the generation profile by assuming 1 W of incident optical power, uniformly distributed along a 300 μm device length, and applied a predefined material absorption coefficient.

### *Electrical field simulation of the waveguide-coupled photodetector structure*

We simulated the internal electric field distribution of the p-i-n junction—the active region of the photodetector—using the CHARGE solver in Ansys Lumerical under a reverse bias of 9 V. The simulation incorporated laterally offset p-type and n-type doping regions, each placed 550 nm from the respective edges of the rib waveguide. We performed a 2D cross-sectional simulation and normalized the device length to 300 μm to ensure consistency with the optical simulation and the generation rate analysis.

*Electrical measurement*

The current-voltage and capacitance-voltage characteristics were measured in the dark using a semiconductor device parameter analyzer (Agilent B1500A).

*Responsivity measurement*

The responsivity of the WG-PD was measured at a wavelength of 1550 nm. A variable optical attenuator (V1550A) and signal generator (DGS3136B) were used to amplitude-modulate the input laser at 70 Hz. The modulated light was split, with one portion directed to a power meter (PM100USB) for real-time monitoring of incident power and the other directed to the WG-PD's grating coupler. The light was coupled into the device at a 10° incidence angle using a cleaved single-mode fiber (SMF-28) after polarization adjustment via a manual polarization controller (FPC651). A trans-impedance amplifier (DLPCA-200) supplied bias to the device and amplified the photocurrent generated by the WG-PD. The amplified signal was subsequently directed to a lock-in amplifier (SRS830), synchronized to the modulation frequency, for precise detection. Output data from the lock-in amplifier were collected and analyzed on a computer. A sketch of the measurement setup is provided in the Supplementary Information (Figure S9). All measurements were performed using TE-polarized light, as the grating couplers are polarization-selective and optimized for TE mode. Simulations confirm that TE is the dominant guided mode in the rib waveguide, ensuring efficient coupling and confinement.

*Noise equivalent power and linear dynamic range measurement*

To accurately determine the NEP and the linear dynamic range of the devices, we measured the device noise spectrum in the frequency domain using a low-noise current preamplifier connected to a spectrum analyzer. A 1550 nm laser, modulated with a 70 Hz square wave, was employed along with a set of neutral density filters to adjust the light intensity incident on the devices, ranging from picowatts to milliwatts. We obtained the NEP by measuring the photodetector current spectral density under the modulated 1550 nm laser across different light intensities, using a measurement bandwidth of 1 Hz. The signal-to-noise ratio (SNR) was then plotted on a double logarithmic scale as a function of the input optical power at 1550 nm. The linear fit of the SNR derived from the frequency-dependent noise current reaches 1, indicating that the signal intensity matches the noise floor as defined by the NEP. This analysis also allows us to determine the linear dynamic range of the detector.

*Bandwidth measurements*

Bandwidth measurements were conducted using a 1550 nm laser, modulated via an electro-optical modulator (IXBlue MXAN-LN-10). The variable frequency modulated light was directed to the WG-PD through a polarization controller (FPC651) and a cleaved optical fiber (SMF-28). A DC power supply, in conjunction with a Bias-T, provided the reverse bias to the WG-PD device. Subsequently, the output photocurrent signal from the device was monitored using a spectrum analyzer (Agilent Tech. PXA N9030A). A sketch of the setup is provided in the Supplementary Information (Figure S11).

## Data availability

The data that support the plots within this paper and other findings of this study are available from the corresponding authors upon request.

## Acknowledgments

We gratefully acknowledge the Ion Beam Center (IBC) and the Nanofabrication Facilities Rossendorf (NanoFaRo) at Helmholtz-Zentrum Dresden-Rossendorf (HZDR) for their assistance with ion implantation and device fabrication. This work was partially funded by the German Research Foundation (DFG) through projects 445049905, 466323332, 498410117, and 528206533.

## Author Contributions

M.S.S. and Y.B. conceived the idea. M.S.S., M.T.C., A.E., R.L., Y.M.G., K.J., and Y.B. designed the device. M.T.C. conducted the simulations under the supervision of K.J. M.S.S., M.C.T., A.E., and R.L. fabricated the devices under the supervision of Y.M.G. M.S.S., S.P., S.Z., and Y.B. designed the implantation and annealing experiments. M.S.S., M.T.C., S.W., and Y.B. developed the characterization setup. M.S.S., M.T.C., S.W., and G.G.P. performed the characterization experiments and analyzed the data under the supervision of Y.B. All authors contributed to the discussion of the experimental results. M.S.S. and Y.B. wrote the manuscript with input from all co-authors. M.H., Y.M.G., K.J., S.Z., and Y.B. supervised and coordinated the entire project.

**Ethics declarations**

Competing interests

The authors declare no competing interests.